\documentclass{aa}  

\usepackage{graphicx}
\usepackage{adjustbox}
\usepackage{amsmath}
\usepackage{color}
\usepackage{txfonts}
\usepackage{hyperref}
\usepackage{amsmath}
\usepackage{xcolor}

\usepackage{media9}

\begin{document}

\title{Streaming instability in neutron star magnetospheres: \\No indication of soliton-like waves}

   \titlerunning{}

   \author{Jan Ben\'{a}\v{c}ek
          \inst{1,2},
          Patricio~A.~Mu\~noz
          \inst{3,2},
          J\"org~B\"uchner
          \inst{2,3}, and
          Axel~Jessner\inst{4}
          }
          
    \authorrunning{Benáček et al.}

    \institute{
        Institute for Physics and Astronomy, University of Potsdam, D-14476 Potsdam, Germany \\ \email{jan.benacek@uni-potsdam.de}
        \and
        Center for Astronomy and Astrophysics, Technical University of Berlin, D-10623 Berlin,
        Germany 
        \and
        Max Planck Institute for Solar System Research, D-37077 G\"ottingen, Germany
        \and
        Max-Planck Institute for Radio Astronomy, D-53121 Bonn, Germany \\
    }

   \date{Received: ; accepted:}

 
  \abstract
   {
       Coherent radiation of pulsars, magnetars, and fast radio bursts could, in theory, be
       interpreted as radiation from solitons and soliton-like waves.
       The solitons are meant to contain a large number of electric charges confined on long time-scales and may radiate strongly by coherent curvature emission.
       However, solitons are also known to undergo a wave collapse, which may cast doubts on the correctness of the soliton radio emission models of neutron stars.
   }
   {
       We investigate the evolution of the caviton type of solitons self-consistently formed by the relativistic streaming instability and compare their apparent stability in 1D calculations with more generic 2D cases, in which the solitons are seen to collapse. 
       Three representative cases of beam Lorentz factors and plasma temperatures are studied to obtain soliton dispersion properties.
   }
   {
        We utilized 1D electrostatic and 2D electromagnetic relativistic particle-in-cell simulations at kinetic microscales. 
   }
   {
       We found that no solitons are generated by the streaming instability in the 2D simulations.
       Only superluminal L-mode (relativistic Langmuir) waves are produced during the saturation of the instability, but these waves have smaller amplitudes than the waves in the 1D simulations. 
       The amplitudes tend to decrease after the instability has saturated, and only waves close to the light line, $\omega = c k$, remain.
        Solitons in the 1D approach are stable for $\gamma_\mathrm{b} \gtrsim 60$, but they disappear for low beam Lorentz factor $\gamma_\mathrm{b} < 6$.
   }
   {
       Our examples show that the superluminal soliton branch that is formed in 1D simulations will not be generated by the relativistic streaming instability when more dimensional degrees of freedom are present. 
       The soliton model can, therefore, not be used to explain the coherent radiation of pulsars, magnetars, and fast radio bursts --- 
      unless one can show that there are alternative plasma mechanisms for the soliton generation.
   }


   \maketitle
%

\section{Introduction} 
\label{sec:intro}
Several coherent radio emission models have been proposed to explain the radiation of pulsars, magnetars, and fast radio bursts \citep{Ruderman1975,Blaskiewicz1991,Kramer2002,Petrova2009,Beskin2018,Xiao2021}.
However, no agreement on a specific mechanism has been reached due to a lack of knowledge on the strongly nonlinear evolution of the relativistic plasma instabilities at kinetic microscales and disparaging theoretical predictions for their coherent emissions \citep{Platts2019,Melrose2020a,Zhang2020}.

The relativistic streaming instability is considered to be a likely cause of coherent radio emission in magnetospheres of radio pulsars \citep{Weatherall1994,Zhang2020} and has also been proposed as the source of radio emissions from magnetar bursts and in magnetar-like models of fast radio bursts (FRBs) \citep{Lu2018,Yang2018,Kumar2020,Yang2020,Lyutikov2021b}.
At distances closer than the light cylinder radius, the streaming instabilities may be produced by pair creation cascades in relativistically hot and highly magnetized open magnetic field line regions above polar caps. \citep{Sturrock1971,Ruderman1975}.
There, the plasma cannot carry the magnetospheric Goldreich--Julian currents parallel to the magnetic field \citep{Goldreich1969}, causing the convective electric field to grow to extreme values of $\sim$10$^{12}$~V\,m$^{-1}$. This field then accelerates electric charges up to relativistic energies, allowing pair creation avalanches, loosely termed ``spark discharges'' to occur.
The created dense pair plasma outflows from the spark region are modulated in time, forming trains of plasma bunches or clouds.
The emitted bunches may expand as they propagate through the magnetosphere, and there are several conceivable reasons why they may induce streaming instabilities:
(1) adjacent bunches may overlap in the phase space, 
(2) the electrons and positrons can have relative drifts, and
(3), the bunches can create instabilities when they propagate through a background plasma \citep{Usov1987,Ursov1988,Asseo1998,Rahaman2020,Manthei2021}.
The streaming instability may also be a source of single pulses from the outer magnetosphere where the magnetospheric reconnection occurs that accelerates the particles to relativistic energies \citep{Cerutti2012,Werner2018,Cerutti2020,Hoshino2020}.

The formation of solitons on kinetic scales in relativistic streaming instabilities has been proposed by \citet{Novikov1984,Sircombe2005,Henri2011}.
Several types of solitons have been considered in the neutron star magnetospheres \citep{Karpman1975,Asseo1998}.
First, soliton-like wave packets called cavitons \citep{Zakharov1972a,Henri2011,Romero2016} may be formed in strong plasma turbulence created by the streaming instability \citep{Eilek2016}. 
In the turbulence, there is an unstable equilibrium between the plasma pressure gradient and the ponderomotive force \citep{Mofiz1988,Mofiz1989}.
This non-equilibrium leads to the creation of plasma cavities (cavitons) filled with a strong electric field and a charge separation inside the so-formed solitons. 
Second, force-free Alfv\'{e}n soliton-like waves were proposed \citep{Mofiz1993,Lyutikov2021a}.
Two types of Alfvén solitons can be formed in the magnetospheres of neutron stars: large-scale waves of the magnetospheric size and waves generated by a firehose type of current-driven instabilities \citep{Pfeiffer2013,Gralla2014,Gralla2015,Lyutikov2020}.
Third, a propagating beam can lead to the Weibel instability in the outer magnetosphere near a current sheet \citep{Dogan2020}.
When the instability saturates, its nonlinear dispersion properties can induce the production of breathers, localized periodic type of solitons, by self-modulation.

Plasma particles can be, in some circumstances, confined in solitons sufficiently long enough to behave collectively, thereby enabling the emission of coherent radiation. 
Such cavity type solitons are assumed to contain a large amount of electric charges. 
The emission power of the coherent emission mechanisms depends on a number of charged particles as $N^2$; hence, the large number of plasma particles associated with the cavitons can be sufficient to explain the observed intensities \citep{Melikidze1980,Melikidze1984,Melikidze2000}. 
It has been proposed that the radiation from solitons could form the strong single radio pulses in the outer magnetosphere of neutron stars \citep{Main2021,Lin2023} or even super-giant pulses \citep{Lyutikov2007} that can be detected over large distances.
The exact evolution, saturation, and the radio emission properties of cavitons in the neutron star magnetospheres are however still being investigation \citep{Eilek2016,Melrose2017a}.

Solitons may radiate via several conceivable emission mechanisms because of their strong nonlinearity and long time stability.
The coherent curvature radiation of cavitons moving along curved magnetic field lines of pulsars is a widely considered radio emission mechanism of neutron stars \citep{Weatherall1991,Melikidze2000,Gil2004,Mitra2009,Melikidze2014,Mitra2017,Rahaman2020b,Horvath2022}.
 
The solitons can also radiate by linear acceleration emission in which the particles undergo parallel acceleration to the magnetic field \citep{Melrose2009a,Melrose2009b}.
However, the power of soliton linear acceleration emission was found to be too low to explain the observed radio power of pulsars \citep{Benacek2023}.

The Alfvén force-free soliton-like waves can be associated with free-electron maser/laser radiation as a model of fast radio bursts \citep{Lyutikov2021b,Lyutikov2021a}.
The Alfvén waves can form wigglers for relativistic particles, and when a particle beam propagates through the wiggler, the plasma particles jiggle in synchronism direction and radiate coherently.
A related model has been proposed by \citep{Dogan2020} where stimulated radio emission is being amplified in klystron-like structures by relativistic bunches created by magnetic reconnection.

The soliton production in neutron star magnetospheres was recently seen in numerical simulations that solve the nonlinear Schr\"odinger equation in one dimension \citep{Lakoba2018}. 
In that approach, the solitons may develop if the \citet{Lighthill1967} condition for the particle distribution is fulfilled, $G q > 0$, where $G$ is the group velocity parameter and $q$ is the cubic nonlinearity parameter.
The solitons were found to be formed with their amplitude growing to infinity because the approach did not consider the feedback of the soliton electric field on the particle distribution.
Using that approach \citet{Rahaman2022}, showed that long-tailed particle distributions, like the Kappa distribution, are more likely to produce the solitons for a wide range of plasma parameters.\\

We have studied the relativistic streaming instability using kinetic particle-in-cell (PIC) simulations that evolve the particles and waves self-consistently in previous publications \citep{Manthei2021,Benacek2021a}. 
These were, however, restricted to one-dimensional (1D) simulations. 
They showed that cavitons can be formed if the plasma inverse temperature is $\rho = mc^2 / k_\mathrm{B}T \geq 1.66 $ (where $m$ is the electron mass, $c$ is the light speed, $k_\mathrm{B}$ is the Boltzmann constant, and $T$ is the thermodynamic temperature), or the Lorentz factor of the beam particles is $\gamma_\mathrm{b} > 40$.
Nonetheless, the required conditions are of greater complexity: The Lorentz factor of the beam can be even $\gamma_\mathrm{b} < 40$ for cavitons to be created when inverse temperatures are higher than $\rho = 1.66$.
We also studied how the streaming instability can be formed in 1D simulations by the overlapping plasma bunches created by spark events \citep{Benacek2021b} and found cavitons with similar properties. 

In this paper, we present the first 2D PIC simulations of the relativistic streaming instability in the inner magnetosphere of a neutron star under the gyro-motion assumption typical for strongly magnetized pair plasma.
We investigate the conditions under which the cavitons may be self-consistently formed or may collapse in 1D and 2D plasmas.
Although soliton is a more general term, we will \textit{use it exclusively from now} on with the understanding that it includes the caviton type of solitary waves.

This paper is structured as follows.
Section~\ref{sec:methods} describes the numerical model and its initial 1D and 2D setups.
In Section~\ref{sec:results}, we present a detailed analysis and comparison of soliton formation in 1D and 2D simulations and the dispersion properties of the resulting electrostatic waves.
We discuss the results and implications on the interpretation of observations by solitons in Section~\ref{sec:discussion} and summarize them in Section~\ref{sec:conclusion}.

\section{Methods} 
\label{sec:methods}
We select three examples S1--S3 cases that we consider to be representative of the range of possible scenarios where initial plasma parameters lead to the formation of the streaming instability and Langmuir turbulence.
The cases follow the three main scenarios of the bunch evolution that are described in the introduction:
(S1) Expanding plasma bunches that start to overlap in the phase space and form the streaming instability.
The streaming instability is characterized by relatively low temperatures and high beam Lorentz factors.
(S2) A streaming instability formed between electrons and positrons inside the bunch, 
Here, the streaming instability is characterized by higher temperatures and lower beam Lorentz factors than in case~(S1).
Furthermore, (S3) represents a plasma bunch propagating through a background plasma.
The streaming instability is characterized by relativistic temperatures and high beam Lorentz factors.

We studied the formation of nonlinear electrostatic solitary waves by the relativistic streaming instabilities in pair pulsar plasma at kinetic microscales using fully-kinetic particle-in-cell (PIC) numerical code ACRONYM \citep{Kilian2012} with the same initial parameters but in separate runs for the 1D and 2D versions of the code.

In all cases, the code is implemented as 1D3V and 2D3V (one or two spatial dimensions and three-dimensional velocity distribution).
Nonetheless, because the particles move strictly along the magnetic field direction, the code is effectively 1D1V and 2D1V (one-dimensional velocity distribution) with the velocity in the direction of the uniform external magnetic field.
We utilize the \citet{Esikperov2001} current-conserving deposition scheme with piecewise quadratic shape (PQS) function of macroparticles.
The PQS shape function described well the plasma dispersion properties of the relativistic plasma, according to our tests.
The code also contains a higher-order shape function ``Weighting with Time-step dependency'' (WT, \citealt{Lu2020}) that we previously used \citep{Manthei2021,Benacek2021a,Benacek2021b}, but according to our tests, it does not provide any better suppression of the numerical Cerenkov radiation artifacts but consumes more computing time.
The simulation setup is summarized in Table~\ref{tab1}.

\begin{table*}[htp]
	\centering
    \caption{Parameters for the 1D and 2D simulations for three simulation cases of the relativistic streaming instability.}
	\begin{tabular}{l|cc|cc|cc}
		\hline \hline
        Simulation case & \multicolumn{2}{c}{S1}\vline & \multicolumn{2}{c}{S2}\vline & \multicolumn{2}{c}{S3} \\
        Simulation type & 1D & 2D & 1D & 2D & 1D & 2D \\
        Simulation name & S1-1D & S1-2D & S2-1D & S2-2D & S3-1D & S3-2D \\
        \hline
        $L \times L$ [$\Delta_\mathrm{x} \times \Delta_\mathrm{y}$] & 8192$\times$1 & 8192$\times$8192 & 8192$\times$1 & 8192$\times$8192 & 8192$\times$1 & 8192$\times$8192 \\
        Time steps [$\Delta t$] & \multicolumn{2}{c}{200\,000} \vline & \multicolumn{2}{c}{223\,000} \vline & \multicolumn{2}{c}{200\,000} \\
        $\Delta x / d_\mathrm{e}$ & \multicolumn{2}{c}{0.1} \vline & \multicolumn{2}{c}{0.2}\vline & \multicolumn{2}{c}{0.1} \\
        $\omega_\mathrm{p} \Delta t$ & \multicolumn{2}{c}{0.05} \vline & \multicolumn{2}{c}{0.045} \vline& \multicolumn{2}{c}{0.05} \\
        $N(x,y,t=0)$ & \multicolumn{2}{c}{Uniform} \vline& \multicolumn{2}{c}{Uniform} \vline& \multicolumn{2}{c}{Uniform}\\
        $N$ [PPC] & \multicolumn{2}{c}{200}\vline & \multicolumn{2}{c}{200} \vline& \multicolumn{2}{c}{200}\\
        $n_1 / n_0$ & \multicolumn{2}{c}{1} \vline& \multicolumn{2}{c}{1} \vline& \multicolumn{2}{c}{1} \\
        $\rho$ & \multicolumn{2}{c}{3}\vline & \multicolumn{2}{c}{2}\vline & \multicolumn{2}{c}{1} \\
        $\gamma_\mathrm{b}$  & \multicolumn{2}{c}{80}\vline & \multicolumn{2}{c}{6.24} \vline& \multicolumn{2}{c}{60} \\
		\hline \hline
	\end{tabular}
    \\ \footnotesize{The simulation domain size in grid cells {$L$, the number of time steps, the grid cell size ($\Delta x$) normalized to the electron skin depth ($d_\mathrm{e}$), the time step ($\Delta t$) in units of plasma frequency ($\omega_\mathrm{p}$), the shape of initial density distribution, the initial number of particles-per-cell $N$~(PPC), beam to background density ratio ($n_1 / n_0$), the inverse temperature $\rho = m_\mathrm{e}c^2 / k_\mathrm{B}T$, and the Lorentz factor of the beam ($\gamma_\mathrm{b}$).}}
	\label{tab1}
\end{table*}

\subsection{Numerical methods}
As the particles move in strongly magnetized plasmas, they lose their perpendicular momenta by synchrotron radiation in less than one simulation time step.
Therefore, we modified the particle pusher to assume that the particle can move only along the magnetic field in a gyro-motion approach.
In the approach, the electric field vector is separated into parallel and perpendicular components to the magnetic field
\begin{equation}
    \vec{E} = \vec{E}_\parallel + \vec{E}_\perp.
\end{equation}
We use the \citet{Vay2011} particle pusher in the gyro-motion (1D velocity space) approximation, which is an energy-conserving ultrarelativistic generalization of the Boris push \citep{Boris1970}.
The pusher considers only the parallel component $\vec{E}_\parallel$ to push the particle, setting the perpendicular velocity of every particle to zero, $v_\perp = 0$.

We assume an initially uniform magnetic field $\vec{B} = 10^{12}$\,G directed along the $x$-axis in the simulation domain.
The gyro-motion approximation in the simulation can be considered valid if most of the perpendicular kinetic energy of a particle is radiated faster than in one simulation time step.
Our checks using computed radiative energy losses found that the condition is fulfilled for all particles in the simulation if the magnetic fields are $\gtrsim 10^{8}$~G.
Assuming a dipolar field with an intensity of $10^{12}$\,G at the star surface, the approximation is valid at least up to a distance of $\sim$22\,$R_\mathrm{NS}$ from the star, where $R_\mathrm{NS}$ is the neutron star radius. 
We note that the applicability of the gyro-motion approximation depends on particle velocity, and the transition between the gyro-motion approach and the requirement for a full 3D velocity calculations is spread over a few orders of magnetic field intensity $\lesssim 10^8$\,G.

To ensure the numerical precision of a perturbation of the magnetic field produced in the simulation, we separate the magnetic field into a constant external component $\vec{B}_0$ and a perturbed component $\delta\vec{B}$
\begin{equation}
    \vec{B}(x, y, t) = \vec{B}_0 + \delta\vec{B}(x, y, t), \quad \vec{B}_0 \gg \delta\vec{B}(x, y, t).
\end{equation}
While the total magnetic field $\vec{B}$ is exploited in the particle pusher, the Maxwell solver uses only the component $\delta\vec{B}(\vec{x},t)$ {as $\nabla \times \vec{B}_0 = \vec{0}$ and $\partial \vec{B}_0/\partial t = \vec{0}$.
The external magnetic field is along the $x$-axis.}

The main numerical challenge in the 2D electromagnetic simulations is the suppression of the numerical Cerenkov radiation appearing as a numerical enhancement of high-frequency waves.
We use the fourth-order M24 finite difference time domain (FDTD) method proposed by \citet{Greenwood2004} for computation on the Yee lattice to suppress these numerical artifacts.
In addition, we combine the field solver with the \citet{Friedman1990} low-pass filtering method with a Friedman parameter $\theta = 0.1$.
This value assures that only the high-frequency waves are suppressed while the waves with frequencies around and at a few multiples of plasma frequency are maintained without significant modification.
We also tested Cole-K\"arkk\"ainen \citep[CK]{Cole1997,Karkkainen2006} field solvers with various parameters, which were stable in 1D electrostatic simulations \citep{Benacek2021a,Benacek2021b}. 
The CK solvers, however, failed to suppress the numerical Cerenkov radiation in electromagnetic 2D simulations while keeping the waves around and above plasma frequency unchanged, even with the applied Friedman filter.

\subsection{Simulation setup of the streaming instabilities}
The plasma is composed of background and beam particles, initially having the same number of particles per cell $n_0 = n_1 = 100$.
We assume both components consist of electrons and positrons, initially in a charge equilibrium in each grid cell.
The particles have an initial Maxwell--J\"uttner velocity distribution function
\citep{Juttner1911}
\begin{equation} \label{eq1}
	g(u) = g_0(u) + g_1(u),
\end{equation}
\begin{equation} \label{eq2}
	g_0(u) = \frac{n_0}{2 K_1(\rho)} \mathrm{e}^{- \rho \gamma},
\end{equation}
\begin{equation} \label{eq3}
	g_1(u) = \frac{1}{\gamma_\mathrm{b}}\frac{n_1}{2 K_1(\rho)} \mathrm{e}^{- \rho \gamma_\mathrm{b} \gamma (1 - \beta \beta_\mathrm{b})},
\end{equation}
where $u/c = \gamma \beta = \gamma v / c$ and $u_\mathrm{b}/c = \gamma_\mathrm{b} \beta_\mathrm{b} = \gamma_\mathrm{b} v_\mathrm{b} /c$
are the four-velocity and velocity of the beam, respectively, both along the $x$-axis and normalized to the speed of light $c$;
$\gamma = (1 - \beta^2)^{-\frac{1}{2}}$ and $\gamma_\mathrm{b} = (1 - \beta_\mathrm{b}^2)^{-\frac{1}{2}}$ are their Lorentz factors,
$K_1$ is the MacDonald function of the first order (modified Bessel function of
the second kind), and
$\rho$ is the inverse temperature.
We assume that there is no net electric charge and no net electric current in the simulations, or in other words, the magnetospheric Goldreich--Julian currents and charges are zero or negligibly small compared to the bunch density.

To the best of our knowledge, there are no specific predictions which plasma parameters produce most of the radio emission. 
We, therefore, follow the estimates of \citep{Arendt2002} for the sparking mechanism and choose the bunches to have an inverse temperature around $\rho \approx 1$ before they evolve. 
This choice appears reasonable in the light of our earlier results for case S1. 
When the bunches overlapped in the phase space, the distribution tended to narrow and evolve to higher inverse temperatures of $\rho \approx 3$ \citep{Benacek2021b}.

The inverse temperature $\rho \approx 2$ in the case S2 was estimated in broad parametric studies, for example, \citet{Rahaman2020a} and \citet{Rahaman2022} to fulfill the Lighthill condition.
The instability then has a high nonlinearity factor in the Lighthill condition and consequently produces solitons.
Similar inverse temperatures can, in addition, be found for highly overlapped plasma bunches.
For S3, we assume a plasma bunch with an inverse temperature $\rho \approx 1$ again but that the bunch propagates through a background plasma of the same temperature.

In simulations S1 and S3, we use the beam Lorentz factors for which the largest linear growth rates are obtained, thereby expected to produce the highest amount of plasma turbulence for the given temperature.
That provides the ``best'' conditions for soliton formation.
Although not rigorously proven, we consider it natural to expect that the largest growth rates of the electrostatic waves will create the strongest instability, the strongest turbulence level, the largest soliton charges, and the largest electromagnetic emission in comparison with the instability with lower linear growth rates.
\citet{Benacek2021a} showed that the largest (1D) growth rates occur for beam Lorentz factors $\gamma_\mathrm{b} \approx 80$ for $\rho=3$ and $\gamma_\mathrm{b} \approx 60$ for $\rho = 1$.
The beam Lorentz factor of $\gamma_\mathrm{b} = 6.24$ for simulations S2 was selected to fulfill the Lighthill condition, as found by \citet{Rahaman2022}. 
That was shown to have a high nonlinearity parameter and to produce solitons in 1D numerical simulations.
For simplicity, we assume the same inverse temperature for the background and beam plasma in all cases.

The grid length of $L = 8192 \Delta_\mathrm{x}$ along the magnetic field is the same for all simulations, where $\Delta_\mathrm{x} = \Delta_\mathrm{y}$ is the grid cell size.
The 2D simulations have 8192$\Delta_\mathrm{y}$ in the perpendicular direction.
The grid length $L$ allows the highest wave number resolution in each dimension up to $\Delta k d_\mathrm{e} \approx 7.7 \times 10^{-3}$ in S1 and S3 and $\Delta k d_\mathrm{e} \approx 3.8 \times 10^{-3}$ in S2.

The simulation time step is chosen as $\omega_\mathrm{p} \Delta t = 0.05$ for simulations S1 and S3 and $\omega_\mathrm{p} \Delta t = 0.045$ for S2.
The highest frequency resolution in the simulation is $\Delta \omega / \omega_\mathrm{p} \approx 1.0 \times 10^{-4}$ in all simulations.
The boundary conditions are periodic in both directions. 
Grid cell size and the time step length differ between simulations because we selected a larger grid cell size and slightly smaller simulation time step in S2 to describe better the produced electrostatic waves in the $\omega-\vec{k}$ space for smaller $\gamma_\mathrm{b}$.

\section{Results} 
\label{sec:results}

\begin{figure*}[ht]
    \includegraphics[width=\textwidth]{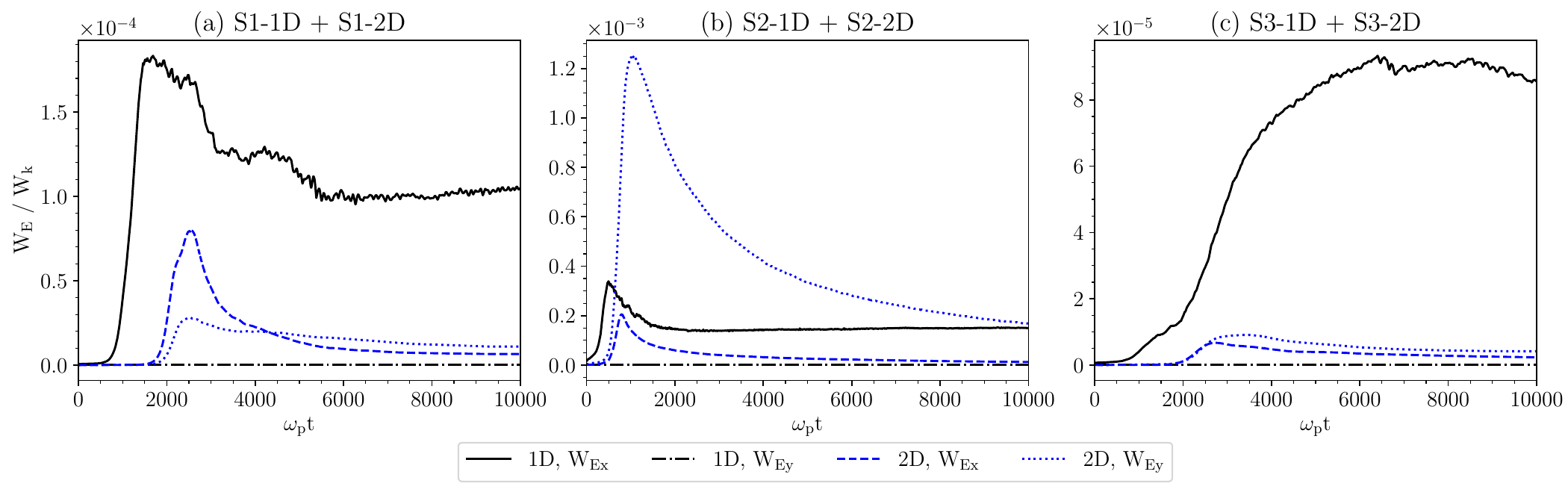}
    \caption{
        Comparison of energy evolution of electric field components parallel $E_\mathrm{x}$ and perpendicular $E_\mathrm{y}$ to the magnetic field between 1D (black lines) and 2D (blue lines) simulations.
        All energies are normalized to the initial kinetic energy.
        The parallel components of the energy ratio $W_\mathrm{x}/W_\mathrm{k}$ are always smaller in 2D than in 1D and decreasing.
        The 2D simulations grow later because they contain a lower initial noise level than the 1D ones at the simulation start.
    }
    \label{fig1}
\end{figure*}

\subsection{Evolution of the electric field}
The evolution of the electric energy $W_\mathrm{E} = W_\mathrm{Ex} + W_\mathrm{Ey}$ in our six simulations is compared in Fig.~\ref{fig1}.
The energy is always normalized to the initial kinetic energy $W_\mathrm{k}$ of the plasma in the given simulation.
The energy components $W_\mathrm{Ex} = \int E_\mathrm{x}^2/8\pi \,d\mathrm{V} = \int w_\mathrm{x} \,d\mathrm{V}$ and $W_\mathrm{y} = \int E_\mathrm{y}^2/8\pi \,d\mathrm{V} = \int w_\mathrm{y} \,d\mathrm{V}$ of the electric field components $E_\mathrm{x}$ and $E_\mathrm{y}$ are then shown for comparison.
The energy ratio evolves in the typical manner of a streaming instability: the instability saturates due to nonlinear effects after an initial exponential growth phase.

The instabilities in the 2D simulations, however, start to grow and saturate later than for 1D simulations because they contain a lower initial noise level, due to a larger number of macro-particles in the Debye volume, from which the instability grows.
This effect is not caused by the different size of the simulations as the electric fields in Fig.~\ref{fig1} are always normalized to the initial kinetic energy of the plasma.
Though the 2D simulations have 8192-times more grid cells, they contain a higher amount of initial kinetic energy by the same factor.
Therefore, the ratio $W_\mathrm{E} / W_\mathrm{k}$ is comparable between 1D and 2D cases.
 
We find that during the whole simulated time interval, the $W_\mathrm{Ex}/W_\mathrm{k}$ energy ratio of the 2D simulations is always smaller than the $W_\mathrm{Ex}/W_\mathrm{k}$ energy ratio for 1D simulations.
Moreover, while the 1D energy evolution stabilizes after the saturation, the 2D simulations show an exponential decay towards a lower equilibrium value.
In the 1D simulations, the $W_\mathrm{Ey}$ energy is always zero, as by definition, there is no electric current nor electric field contribution to this component in the 1D1V approximation.
The $W_\mathrm{Ey}$ energy is plotted only for clarity.
However, the $W_\mathrm{Ey}$ energy is nonzero in 2D simulations.
In simulation S1-2D, that component saturates at lower energy, 
but it later exceeds the $W_\mathrm{Ex}$ energy, which decreases faster.
In simulations S2-2D and S3-2D, the $W_\mathrm{Ey}$ energy is the same or greater than the $W_\mathrm{Ex}$ one.
Both components grow and saturate together in 2D simulations; hence, their evolution can be considered as being strongly connected.

\begin{table*}[!htp]
	\centering
    \caption{Growth rates, dissipation rates, and asymptotic energy ratios for all simulations.
        Because the $W_\mathrm{Ey} = 0$ for the 1D simulations, the corresponding parameters are not estimated.
    Also, S1-1D and S3-1D do not exponentially decrease.}
	\begin{tabular}{l|cc|cc}
		\hline \hline
         & \multicolumn{2}{c}{Growth rates} \vline & \multicolumn{2}{c}{Dissipation rates and asymptotic energies} \\
         \hline
        Simulation  & $\Gamma(W_\mathrm{Ex})/\omega_\mathrm{p}$ & $\Gamma(W_\mathrm{Ey})/\omega_\mathrm{p}$ & $\delta(W_\mathrm{Ex})/\omega_\mathrm{p}$ &  $\delta(W_\mathrm{Ey})/\omega_\mathrm{p}$ \\
         &  &  & $W_\mathrm{asymp}(W_\mathrm{Ex})$ & $W_\mathrm{asymp}(W_\mathrm{Ey})$ \\
        \hline
        S1-1D & $(8.99 \pm 0.01) \times 10^{-3}$ & - & - & - \\
         &  &   & $(1.0 \pm 0.1) \times 10^{-4}$ & - \\
        \hline
        S1-2D & $(8.48 \pm 0.06) \times 10^{-3}$ & $(7.62 \pm 0.02) \times 10^{-3}$ & $(2.24 \pm 0.03) \times 10^{-3}$ & $(1.06 \pm 0.01) \times 10^{-3}$ \\
         &  &  & $(2.0 \pm 0.2) \times 10^{-5}$ & $(1.7 \pm 0.03) \times 10^{-5}$ \\
        \hline
        S2-1D & $(1.12 \pm 0.05) \times 10^{-2}$ & - & $(1.2 \pm 0.2) \times 10^{-3}$ & - \\
         &  &  & $(1.4 \pm 0.1) \times 10^{-4}$ & - \\
        \hline
        S2-2D & $(1.71 \pm 0.02) \times 10^{-2}$ & $(1.49 \pm 0.01) \times 10^{-2}$ & $(2.45 \pm 0.03)\times10^{-3}$ & $(6.31 \pm 0.05) \times 10^{-4}$ \\
         &  &  & $(5.38 \pm 0.05) \times 10^{-5}$ & $(2.5 \pm 0.03) \times 10^{-4}$ \\
        \hline
        S3-1D & $(4.38 \pm 0.02) \times 10^{-3}$ & - & - & - \\
              &  &   & $(4.5 \pm 0.9) \times 10^{-5}$ & - \\
        \hline
        S3-2D & $(5.76 \pm 0.08) \times 10^{-3}$ & $(4.82 \pm 0.07) \times 10^{-3}$ & $(4.04 \pm 0.01) \times 10^{-4}$ & $(5.15 \pm 0.07) \times 10^{-4}$ \\
         &  &  & $(2.2 \pm 0.3) \times 10^{-6}$ & $(3.9 \pm 0.2) \times 10^{-6}$ \\
		\hline \hline
	\end{tabular}
	\label{tab2}
\end{table*}

We estimate the growth rates $\Gamma$ of all simulations and dissipation/decrease rates $\delta$ of the 2D simulations separately for both amplitudes of electric field components in Table~\ref{tab2}.
The rates are obtained by an exponential fit of the energy ratio evolution profile, and they are normalized to the plasma frequency $\omega_\mathrm{p}$.
The time windows of the highest and lowest rates are selected for the growth and decrease, respectively, in each simulation.
Because the $E_\mathrm{y}$ component equals zero for all 1D simulations, only the parameters for the $E_\mathrm{x}$ component are estimated.
The growth rates of the energy ratios are assumed in the form of $W_\mathrm{E}/W_\mathrm{k} \approx  W_0 \mathrm{e}^{2 \Gamma t} + W_\mathrm{1}$, where $W_0$, $W_1$, and $\Gamma$ are the parameters for the fit.
The dissipation rates are assumed similarly $W_\mathrm{E}/W_\mathrm{k} \approx W_0\mathrm{e}^{-2 \delta t} + W_\mathrm{asymp}$, where $W_\mathrm{asymp}$ is an asymptotic value to which the energy ratio approach after the saturation stage.

The growth rate values of $W_\mathrm{Ex}/W_\mathrm{k}$ ratio are comparable between 1D and 2D simulations.
The simulation S2-2D has the highest growth rate in both $W_\mathrm{Ex}$ and $W_\mathrm{Ey}$.
While the simulations S2-2D has highest dissipation rate in $W_\mathrm{Ex}/W_\mathrm{k}$ ratio, S1-2S features, in contrast, the highest dissipation rate in $W_\mathrm{Ey}/\varepsilon_\mathrm{k}$ ratio.
The asymptotic energies of both components of 2D simulations are significantly lower than their saturation energies.

\begin{figure*}[htp]
    \centering
    \includegraphics[width=\textwidth]{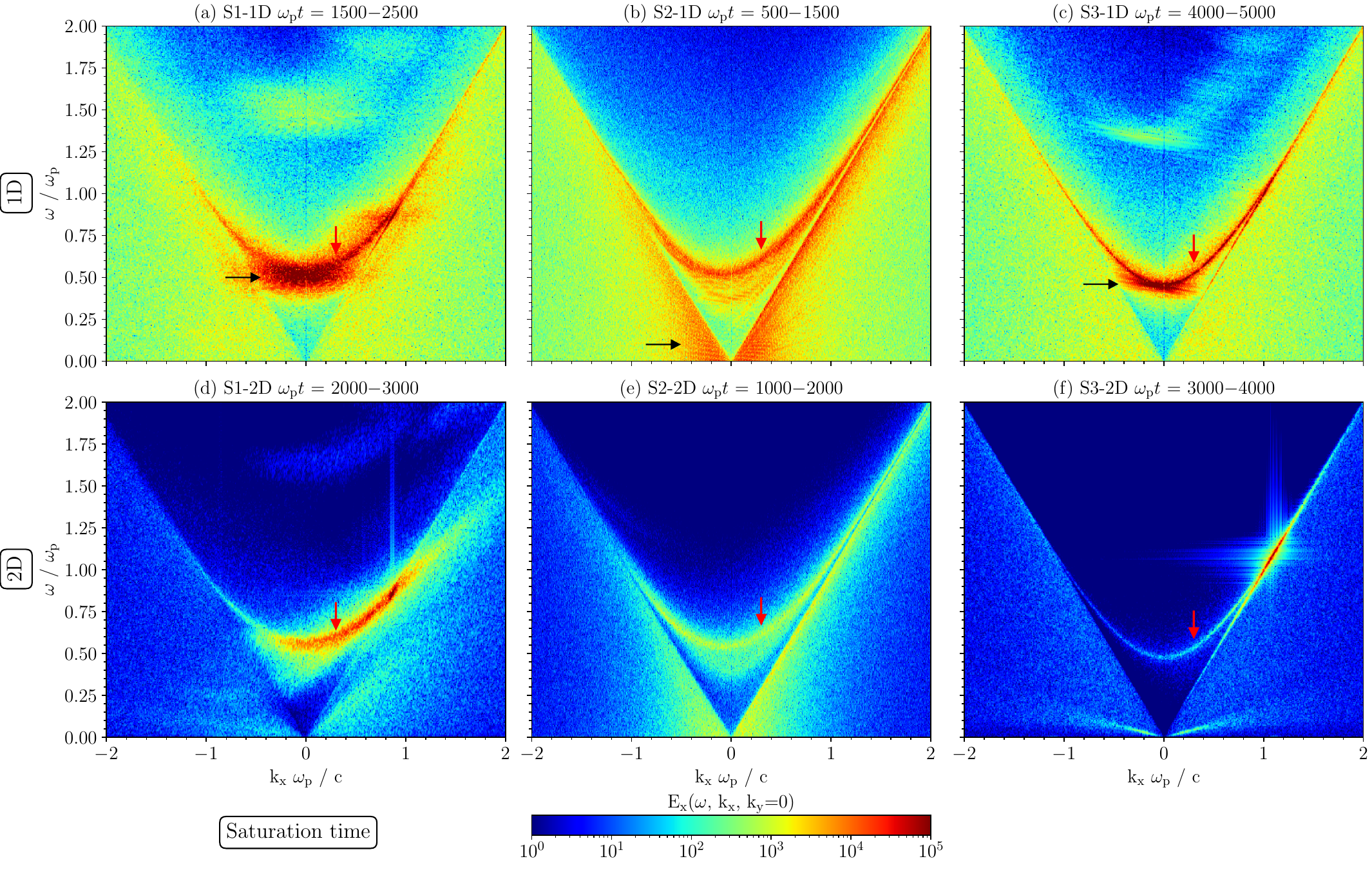}
    \caption{
        Dispersion diagrams of the electrostatic waves along the magnetic field around the saturation time.
        Comparison between 1D (top row) and 2D (bottom row) simulations with the same plasma parameters.
        The solitons dispersion branches are denoted by black arrows and the superluminal L-mode waves by red arrows.
        The superluminal waves appear in 1D and 2D, while the soliton branch appears only in 1D.
    }
    \label{fig2}
\end{figure*}

\begin{figure*}[htp]
    \centering
    \includegraphics[width=\textwidth]{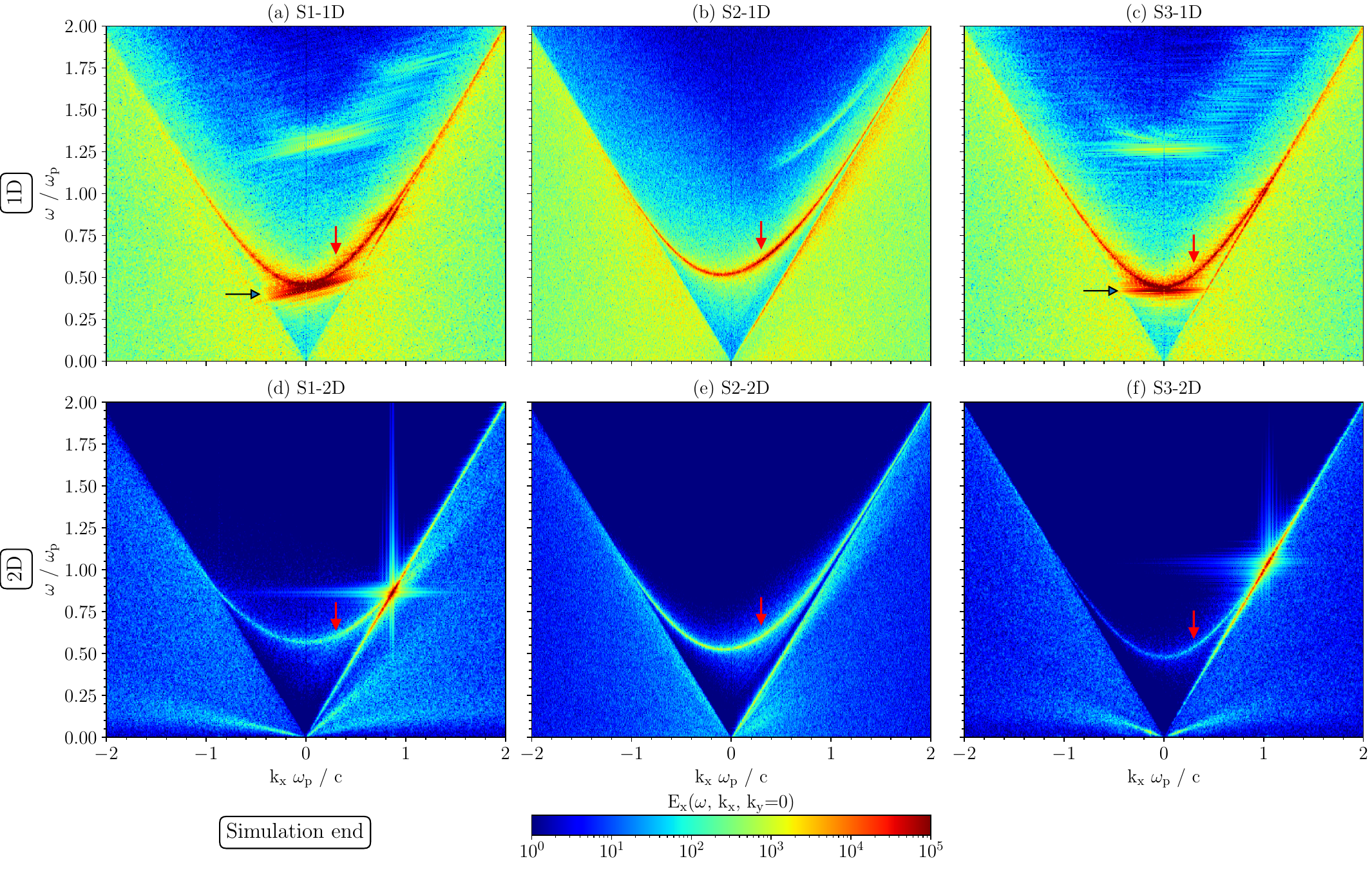}
    \caption{
        The same as Fig.\ref{fig2} but at the end of simulations in a time interval $\omega_\mathrm{p}t =$\,9000--10\,000.
        Solitons do not appear in simulations (b) and (d)--(f).
        While the most intensive waves in 1D are the solitons and superluminal L-mode waves close to $k_\parallel \approx 0$, in 2D, the most intensive waves are close to the light line, and solitons are not present.
    }
    \label{fig3}
\end{figure*}

\subsection{Identification of wave modes}
Figs.~\ref{fig2} and \ref{fig3} show dispersion diagrams of the electrostatic waves.
Specifically, the $E_\mathrm{x}$ component along the magnetic field line is chosen because this component is common for the 1D and 2D simulations.

The limited storage capacity of the SuperMUC-NG cluster did not allow us to store the field data for the whole simulation time. 
We have, therefore, chosen a time interval of $\omega_\mathrm{p} \delta t = 1000$ based on the saturation time for the dispersion diagrams, and the similar time intervals were used in Fig.~\ref{fig2} for the saturation regime and in Fig.~\ref{fig3} for the end of the simulation at $\omega_\mathrm{p} t = 9000$--10\,000.

All subplots of Figs.~\ref{fig2} and \ref{fig3} show longitudinal L-mode electrostatic waves indicated by red arrows.
In contrast, we associated the solitons with horizontal branches of the superluminal electrostatic waves.
The superluminal L-mode waves can be obtained as solutions of the linearized relativistic dispersion permittivity tensor in the gyro-motion approximation, $\Lambda_\mathrm{33} = 0$ \citep{Rafat2019a}. 
The soliton branch is, however, a \textit{nonlinear} solution of the dispersion tensor. 
The soliton waves form straight, almost horizontal lines in the $\omega - k$ frequency--wave number space for $k c/\omega_\mathrm{p} \lesssim 1$ and $\omega \lesssim \omega_\mathrm{p}$ (where $\omega_\mathrm{p}$ is the plasma frequency), thereby not following the Langmuir dispersion branch.
A more detailed analysis of the visible horizontal dispersion features based on the inverse Fourier transformation to real space of the horizontal branch area in the $\omega - k$ domain showed that they correspond to stable, soliton-like waves \citep{Benacek2021a}.
Furthermore, we therefore identify the visible horizontal structures as the \textit{soliton dispersion branch} in the rest of the paper. 
Because the dispersion branch is horizontal in the soliton reference frame, all the waves associated with the soliton oscillate with the same frequency but with different $k$ in that frame.
Their group velocities are also the same as their dispersion branch, which has a constant slope.
Hence, the soliton branch can be described by a dispersion relation $\omega(k) = \omega_\mathrm{s}$, where $\omega_\mathrm{s}$ is the soliton oscillation frequency.
The soliton dispersion branches are indicated by black arrows in Figs.~\ref{fig2} and \ref{fig3}.

\subsection{Analysis of the simulation results and the evolution of plasma waves in different scenarios}
In simulations S1-1D and S3-1D, almost horizontal straight branches can be identified closely below the cutoff frequency of the superluminal L-mode waves. 
That, and the broad horizontal extent in the $\omega - k$ diagram, indicates that there have to be correspondingly narrow spatial features with very low group velocity that are repetitive and spiky in time, which is typical for plasma solitons.

The simulation S2-1D shows the soliton branches during the saturation in Fig.~\ref{fig2} but not at the simulation ends in Fig.~\ref{fig3}.
The soliton branches in S2-1D are different from S1-1D and S3-1D; mostly subluminal and located more below the L-mode cutoff frequency, typically at lower frequencies around $\omega = (0-0.3)\omega_\mathrm{p}$.
We found that the main factor influencing the difference is the Lorentz factor of the beam, which is significantly lower in S2 than in the other two cases. 
Moreover, compared with the 1D simulations, the 2D simulations show no evidence of a soliton branch neither at saturation time nor at the simulation end.
Most energy at the end of simulations S1-2D and S3-2D is present at the L-mode branches in a region of the branch close to the light line $\omega = c k$.
In the simulation S2-2D, approximately the same amount of energy is present in the superluminal and the subluminal L-mode branches.
However, we find that the superluminal waves continue to dissipate slowly at the simulation end.

\begin{figure*}[htp]
    \centering
    \includegraphics[width=\textwidth]{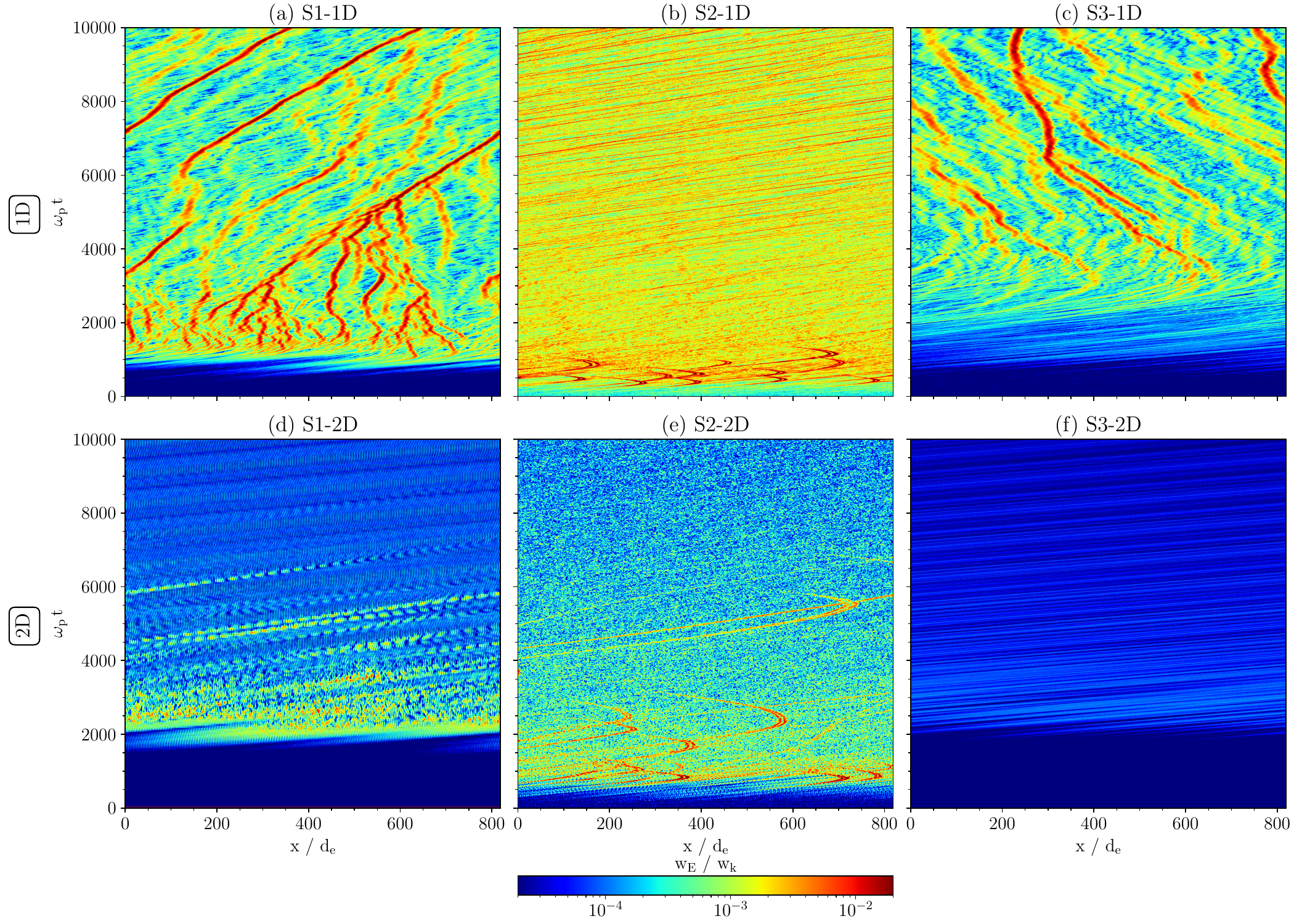}
    \caption{
        Evolution of the electrostatic energy density in simulations normalized to the initial kinetic energy density.
        Although in S1-1D and S3-1D, the solitons survive until the simulations end, in S2-1D, the solitons disappear before $\omega_\mathrm{p}t = 2000$.
        {No solitons appear in 2D simulations.}
    }
    \label{fig4}
\end{figure*}

\begin{figure*}[htp]
    \centering
    \includegraphics[width=\textwidth]{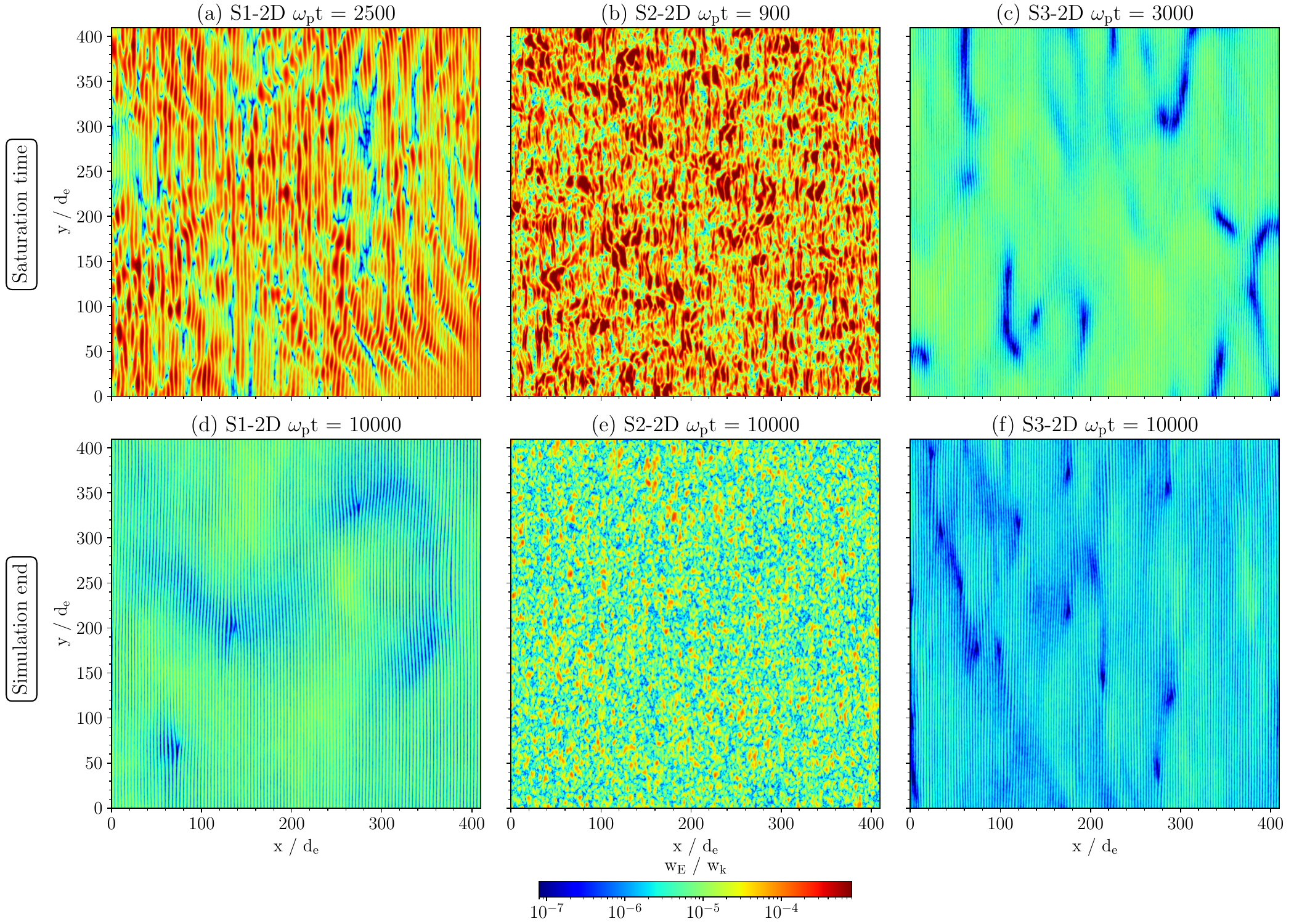}
    \caption{
        The electrostatic energy density of the 2D simulations normalized to the initial kinetic energy density in the $x-y$ plane during the saturation times (top row) and at the simulation ends (bottom row). 
        The evolution is available as an online movie.
        Only a simulation subdomain is selected to show the structures formed by the waves.
        The energy evolution is available as an \textit{online movie}.
    }
    \label{fig5}
\end{figure*}

In Figs.~\ref{fig4} and \ref{fig5}, we analyze the spatial profiles of the electrostatic energy density $w_\mathrm{E} = |E|^2/8\pi$ normalized to the initial kinetic energy density $w_\mathrm{k}$ separately for 1D and 2D simulations.
The 2D simulations are selected as a cut along $x$ at $y = L/2$ axis, where $L$ is the size of the simulation domain, and bicubic interpolation is applied in all plots.
In 1D simulations, electric fields were stored on disks every 20th time step, but in 2D simulations, we stored the fields only every 200th time step in S2-2D and S3-2D and every 1000th time step in S1-2D because of storage limitations. 
Therefore, the structures in Fig.~\ref{fig4}(d) are worse resolved than other subfigures of the plot.
The soliton waves appear during the instability saturation as the most intensive red regions in the S1-1D and S3-1D simulations in Fig.~\ref{fig4}, and we associated these waves with the soliton branches in Figs.~\ref{fig2} and \ref{fig3}.
Although the solitons survive until the end of the simulations in S1-1D and S3-1D, they are not very visible in Fig.~\ref{fig4}(b) because their intensity is smaller than the intensity of the superluminal waves, which can be seen from Fig.~\ref{fig2}(b).
The superluminal waves decrease quickly in S2-1D and S2-2D after their saturation approximately in $\omega_\mathrm{p}t \sim 1000$.
In the S2-1D simulation, only the L-mode waves remain, seen as almost horizontal orange-red lines moving with group velocities close to the light speed.
The simulations S1-2D and S3-2D also show the rapid decrease of electrostatic energy.

We plot the electrostatic energy density in 2D simulations in Fig.~\ref{fig5} during the saturation times and at the simulation ends.
One quarter of the simulation domain is shown to better resolve the electric field structures in the plot.
During the saturation time, more irregularly red-like shaped regions are present in Fig.\ref{fig2}.
We found the regions associated with the superluminal L-mode waves in dispersion diagrams of Fig.~\ref{fig2}(d) and (e).
The vertical green--yellow stripes along $y$-axis that are narrow in $x$ in Figs.~\ref{fig5}(c), (d) and (f) as well as the small green-yellow regions in Fig.~\ref{fig5}(e) are associated with the L-mode waves close to the light line.

\begin{figure*}[htp]
    \centering
    \includegraphics[width=\textwidth]{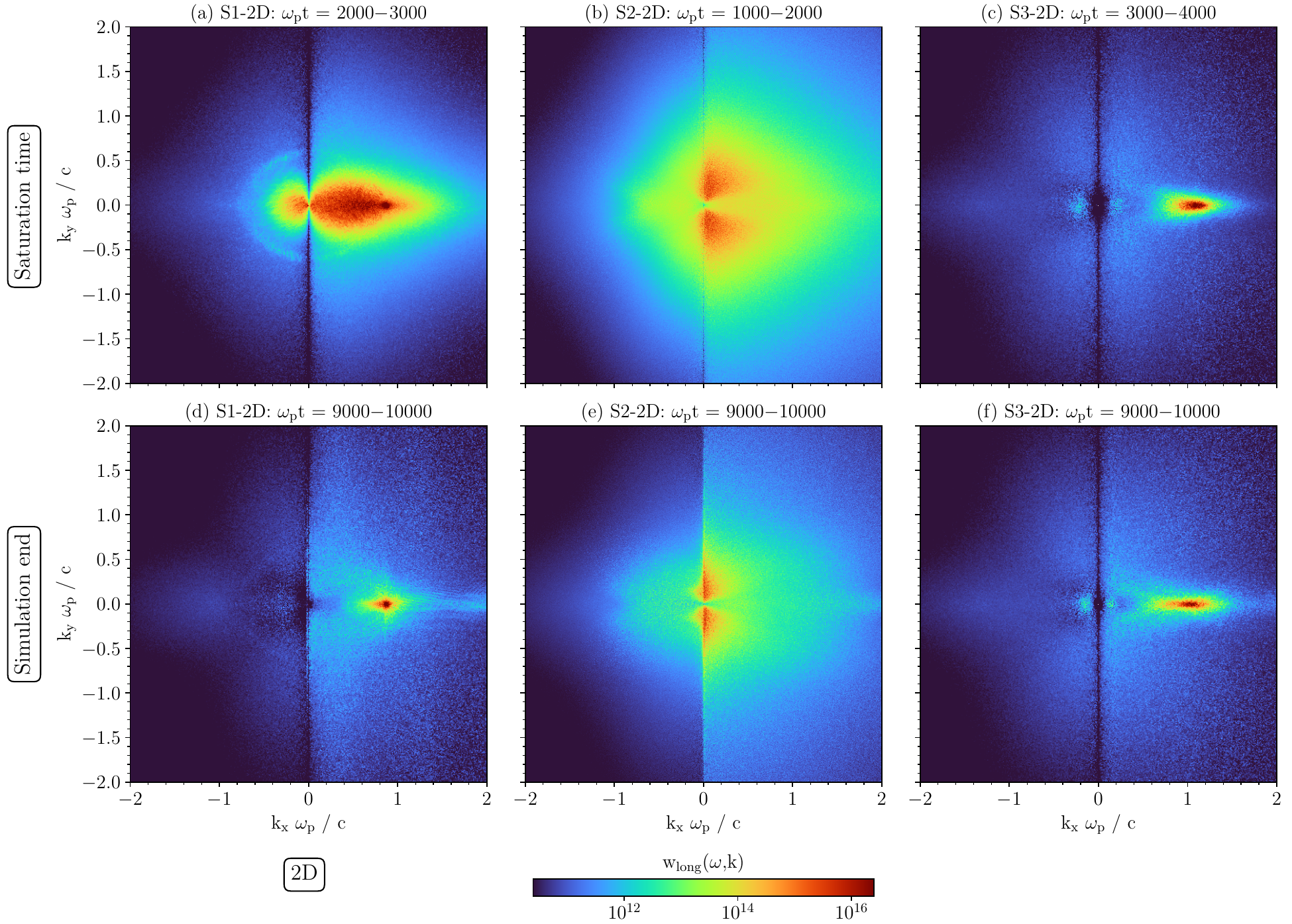}
    \caption{
        The energy density of longitudinal electrostatic waves of 2D simulations in the wave vector space during the saturation time intervals (top row) and at the simulation ends (bottom row).
        All analyzed intervals are $\omega_\mathrm{p}t_\mathrm{a} = 1000$ long.
        The comparison between rows shows that the longitudinal electrostatic waves close to $|\vec{k}| = 0$ decrease in time.
    }
    \label{fig6}
\end{figure*}

Figure~\ref{fig6} compares the wave vector space of longitudinal electrostatic waves in the 2D simulations at saturation time and simulation end.
The longitudinal electric component for a grid cell ($i,j$) in the $k_\mathrm{x}-k_\mathrm{y}$ space is obtained as $E_\mathrm{long}(i,j) = k_\mathrm{x} E_\mathrm{x}(i,j) + k_\mathrm{y} E_\mathrm{y}(i,j)$ for each wave vector $\vec{k} = (k_\mathrm{x}, k_\mathrm{y})$.
The wave energy density is calculated as $w_\mathrm{long}(k_\mathrm{x}, k_\mathrm{y}, \omega) = \frac{1}{8\pi}|E_\mathrm{long}(k_\mathrm{x}, k_\mathrm{y}, \omega)|^2$ where $E_\mathrm{long}$ is in Gaussian CGS units of the simulation. 
The waves are obtained in $\omega_\mathrm{p}t = 1000$ long intervals and summed over frequencies in a range $\omega = (0-2)\omega_\mathrm{p}$.
While in S1-2D and S2-2D simulations, the wave profiles significantly change after the instability saturation, only a small narrowing of the waves appears along the $x$-axis in S3-2D.
Moreover, in S1-2D and S3-2D, the waves have directions mostly along the magnetic field at the simulation end.
In S2-2D, the most intense longitudinal waves are directed perpendicular to the magnetic field.
These longitudinal waves along the $y$ axis are associated with the intensive $E_\mathrm{y}$ component (blue dotted line) in Fig.~\ref{fig1}(b).
Note that the longitudinal waves propagating in the perpendicular direction to the magnetic field are described by the $E_\mathrm{y}$ component of the electric field in our configuration.
A deeper analysis in the $\omega - k_\mathrm{x} - k_\mathrm{y}$ space, revealed that these perpendicular waves have frequencies $\omega < 0.01\omega_\mathrm{p}$ for $|k_\mathrm{y} c /\omega_\mathrm{p}| \gtrsim 0.5$ and $\omega < 0.05\omega_\mathrm{p}$ for $|k_\mathrm{y} c /\omega_\mathrm{p}| \lesssim 0.5$.

\subsection{Evolution of particle distributions}

\begin{figure*}[!htp]
    \centering
    \includegraphics[width=\textwidth]{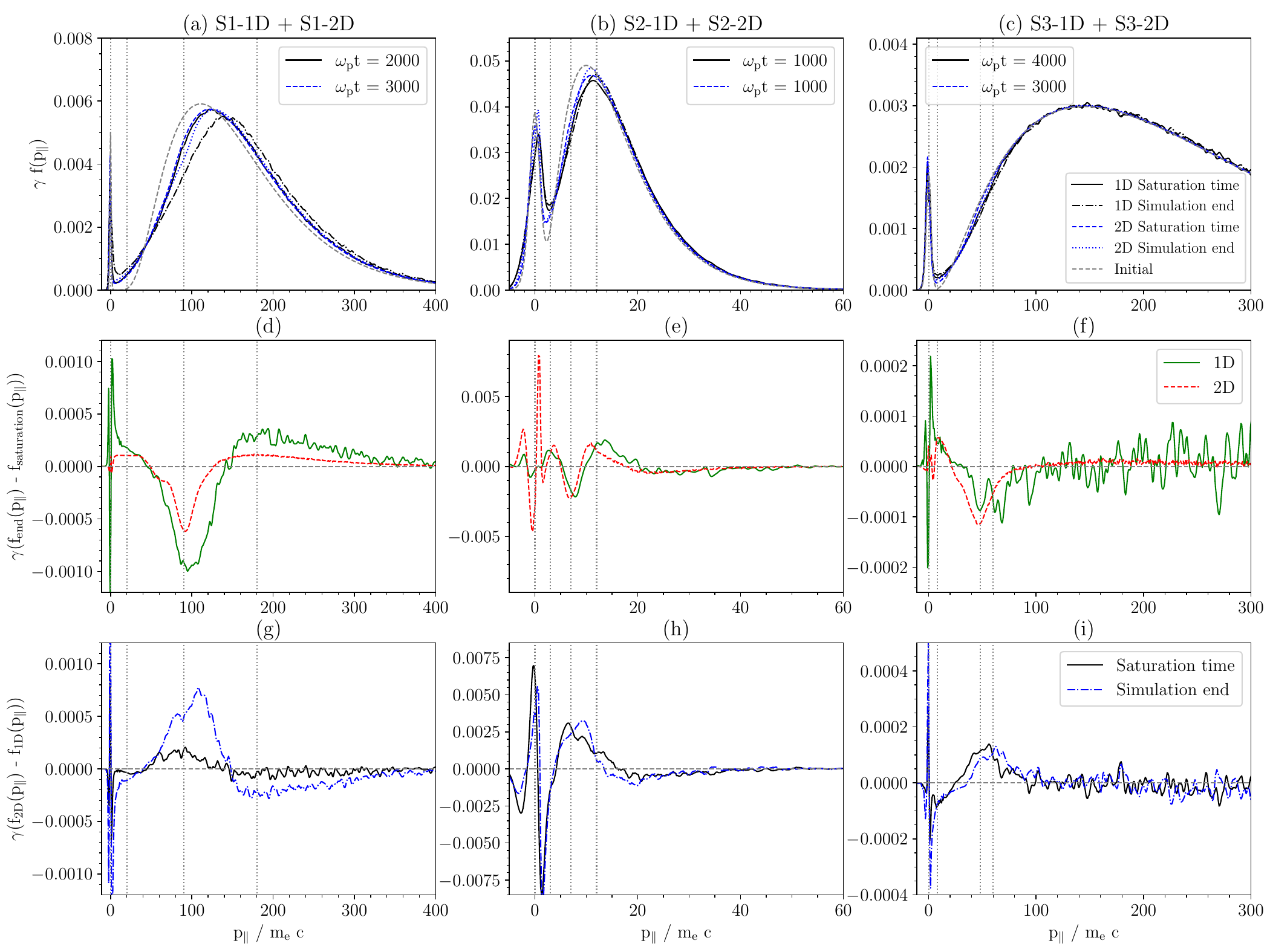}
    \caption{
        Comparison of particle distributions as functions of the particle momenta.
        All electrons and positrons are used from the simulations.
        The distributions for 1D and 2D simulations at the simulation start, instability saturation, and the simulation end (top row).
        Comparison of the distribution difference between the simulation end and saturation time for 1D and 2D simulations (middle row). 
        Comparison of the distribution difference between 1D and 2D simulations for the saturation time and simulation end (bottom row).
        The grey vertical dotted lines show selected positions of momenta where the distributions change significantly.
        The grey dashed lines show the initial distributions (top row) and a zero distribution difference (middle and bottom rows).
    }
    \label{fig7}
\end{figure*}

Figure~\ref{fig7} shows the particle distributions $\gamma f(p_\parallel)$, where $f(p_\parallel)$ is the particle momentum distribution, at the start of the simulation, at instability saturation time, and at the end of the simulation for 1D and 2D simulations.
All simulation macro-particles (electrons and positrons) are included, and local variations in the distributions are averaged.
The distributions are normalized as $\int_{-\infty}^{\infty} \gamma(p_\parallel)f(p_\parallel) \mathrm{d}p_\parallel = 1$.
Only the momentum intervals with significant evolution of the distributions are shown.
The beams in 1D simulations have a relatively small number of particles, and their distributions contain a high level of noise; the beam parts of these distributions were, therefore, smoothed using the Gaussian filter with a width of four data points.

In all simulations, the initial distributions relax during the linear evolution to the saturation time, generating electrostatic waves (Fig.~\ref{fig1}) and thereby closing the gap between the beam and background populations (Fig.~\ref{fig7}(a--c)). 
Later on, during the saturated stage of the instability, the distribution functions evolve slowly to an asymptotic state with profiles in the velocity space between the initial background and beam populations.
We find that the saturation and asymptotic states of the 2D distributions have steeper slopes in comparison with the 1D cases. 
This implies that more kinetic energy is converted to waves in the 1D simulations than in the 2D simulations, which agrees with the behavior seen in Fig.~\ref{fig1}, excepting, however, the $W_\mathrm{Ey}$ component of S2-2D, which may not be directly related to the average velocity distribution along the magnetic field lines but rather to spatial density fluctuations perpendicular to the magnetic field.
Though the 1D and 2D distributions are similar at saturation times, their small differences are significant enough to explain the differences in the evolution of the electrostatic wave energy.
The change is caused by a redistribution of the kinetic energy over the distribution function towards smaller momentum gradients of the distributions in regions with positive, $p_\parallel \partial f / \partial p_\parallel$, resulting in higher plasma stability.

Figure~\ref{fig7}(d--f) shows the differences in distributions between the saturation time and the simulation end.
There are regions in momenta where the distributions decrease and regions where they increase, respectively.
The largest values of the differences are for the background plasma momenta $|p_\parallel / (m_\mathrm{e} c) | \lesssim 1$; however, these differences are not associated with large energy changes because the differences are restricted to very narrow momentum space intervals.
Between saturation time and simulation end, the distributions evolve mainly in regions where the particle distributions 
(1) increase for momenta between the background and beam distributions, 
(2) decrease where the distribution has a positive gradient, and 
(3) increase in the beam tail where the distribution has a negative gradient.
The momentum distribution gradients become flatter in time in all cases.

The distributions differ between the 1D and 2D simulations as shown in Fig.~\ref{fig7}(g--i).
The largest differences are for beam momenta where the distributions have positive gradients.
The distribution functions of 2D simulations have systematically higher values than in 1D simulations in this region.
For higher momenta where the distributions have negative distribution gradients for beam particles, the distribution functions of 2D simulations have smaller or similar values.
The differences between the background distributions of 1D and 2D simulations are negligible in S1 and S3 simulations because the distributions are very narrow in momentum space.
In the S2 simulations, the negative momentum distribution tail $p_\parallel / (m_\mathrm{e} c) \lesssim -3$ has higher values for 1D simulation, for momenta $p_\parallel / (m_\mathrm{e} c) \approx -3$ to $1$ is higher for 2D simulations, and for momenta $p_\parallel / (m_\mathrm{e} c) \approx 1-4$.
As a result, the 1D distributions have systematically smaller momentum gradients than the 2D ones.

\section{Discussion}
\label{sec:discussion}
We studied the formation and evolution of cavitons, soliton-like waves, as possible sources of coherent pulsar radio emissions by utilizing 1D and 2D particle-in-cell (PIC) simulations of the relativistic streaming instability in strongly magnetized pair plasma of neutron star magnetospheres.

Focusing on the instability evolution during and after saturation, we found that the solitons, known to form a horizontal dispersion branch in the 1D simulations, do not appear in 2D simulations because of the additional degree of freedom perpendicular to the magnetic field in the plasma.
Only strong superluminal L-mode waves on the relativistic Langmuir-like dispersion branches appear in 2D simulations during the saturation of the instability.
These waves decrease in time-scales of a few hundreds or thousands of plasma periods.
Moreover, in one of the 1D simulations, the solitons amplitudes were also found to decrease, and the solitons collapse after some time.
Solitons being already of a somewhat precarious existence in 1D simulations cannot be considered a realistic contender for radio emission in strongly magnetized pulsar plasma based solely on the 1D approach.
Soliton-like waves do not appear at all in the 2D simulations, a finding that agrees with the already described soliton collapse for unmagnetized or weakly magnetized kinetic plasmas, where the solitons may initially appear but collapse quickly when their wavelengths increase during their evolution \citep{Zakharov1972b,Robinson1997}.

In contrast to the 1-D soliton stability, there is no stable solution of the nonlinear Schrödinger equation in more than one dimension for space and laboratory plasmas \citep[and references therein]{Robinson1997}.
The Langmuir solitons in electron--ion plasma were shown to undergo collapse and dissipation when two-dimensional (2D) plasmas are considered, which was shown by solving the set of Zakharov equations and also confirmed by numerical PIC simulations for weakly magnetized plasmas \citep{Newman1990}.
The main difference between our and the classical approach based on solving of the Schrödinger equation and Zakharov equations is that the PIC simulations describe the particle--wave interactions self-consistently.
Specifically, the PIC simulations consider the evolution of the particle distribution as a function of space and time.
And exactly these changes of the distribution function may be the nonlinear effects that suppress the formation of the solitons in the 2D approach.
We showed, in addition, that the nonlinear soliton dispersion forms a straight horizontal branch in the frequency--wave number space in the 1D approach. 
However, the analytical description is based on a soliton dispersion that follows the quasilinear dispersion relation of Langmuir- or L-modes. However, that dispersion branch does not contain the superluminal or subluminal L-modes that are also present.

 A soliton being a plasma structure affecting the electric field and the particle distribution function can only be influenced by electric fields along the magnetic field in 1D simulations. The physical reason for the absence of solitons in the 2D simulations is connected with the additional plasma effects in directions perpendicular to the magnetic field.
In the soliton reference frame, the soliton dispersion branch is horizontal, which agrees with expected soliton stability as all waves forming the soliton oscillate with the same frequency and have the same group velocity.
The solitons are assumed infinitely large in the $y$ direction in 1D, but their true perpendicular size in 2D or 3D remained as an open question.
 In the 2D model, particles located perpendicular to the locally enhanced electric field of the soliton progenitor are free to also interact with their surroundings in the perpendicular direction to the magnetic field and the electric field of the wave can spread out and decrease.
These superluminal L-mode waves have been observed to decrease faster in S1-2D and S3-2D than in S2-2D, where the perpendicular size of the waves is smaller.
Figure~\ref{fig5} must be seen as supporting evidence that the effects of large perpendicular size contribute significantly to the wave decay.

The particle distributions simultaneously stabilize when the electrostatic wave energy in the 2D simulations decreases after saturation, see Figs.~\ref{fig1} and \ref{fig7}.
The momentum gradients of averaged particle distributions, $p_\parallel \partial f / \partial p_\parallel$, decrease in time, and the instability growth rates approach zero in all 1D and 2D simulations.
The 1D distributions have systematically smaller gradients of the distributions at the saturation time than the 2D ones, implying higher energy conversion in 1D simulations than in 2D ones.

The conversion of the electrostatic energy after the instability saturation occurs during the nonlinear evolution phase of the instability.
There are two reasons why it is difficult to find direct evidence of the electrostatic energy conversion in the average particle distribution profiles.
First, although the parallel (to the magnetic field) subluminal component of electric waves can be converted into  particle kinetic energy, the energy is small compared to the total energy that is simultaneously being redistributed between various parts of the particle distribution.
Second, the presented particle distributions are spatially averaged, but part of the electrostatic energy is associated with the local density fluctuations that are suppressed by such averaging.
The amplitudes of these density fluctuations also decrease after the instability saturation, and that also decreases the amplitudes of the parallel superluminal electrostatic and perpendicular electrostatic waves.
There cannot be direct energy conversion of the parallel superluminal waves because wave--particle interaction can only occur for subluminal wave speeds.
Direct conversion of the perpendicular wave energy to particle motions is also ruled out because the particles move strictly along the magnetic field, and the electric field is perpendicular to their velocity vector.
Both types of waves are instead converted nonlinearly by the fluctuations distributed along and across the magnetic field.

The Lorentz factor of the beam influences the directivity of the longitudinal waves at the end of the 2D simulations (Fig.~\ref{fig6}).
The longitudinal waves are directed closely along the magnetic field for high values $\gamma_\mathrm{b} \gtrsim 60$, but low frequency waves are enhanced mainly in the $y$ direction for lower $\gamma_\mathrm{b} \approx 6$ (S2-2D).
These waves have higher intensities than the waves along the magnetic field in the same simulation, and the wave amplitudes slowly decrease.

We found that the growth rates for L-waves do not differ strongly between 1D and 2D simulations.
As was shown by \citet{Manthei2021}, the growth rates obtained from the 1D PIC simulations are comparable with those obtained from a linearized analytical 1D approach.
The higher growth rates of $W_\mathrm{Ex}$ compared to $W_\mathrm{Ey}$ are probably caused by 
nonlinear coupling between both components. 
The decay rates are, however, lower than the growth rates in all three 2D simulations.

The 1D and 2D comparison indicates that the gradients of $\partial f(x,y, \vec{v})/\partial y \neq 0$ and $\partial \vec{E}(x,y) / \partial y \neq 0$ may be important for the suppression of soliton formation.
 These gradients are equal to zero in the 1D approach, $\partial f(x,y)/\partial y = 0$ and $\partial \vec{E} / \partial y = 0$, but are allowed to vary in 2D.

Some laboratory results appear to show the formation of soliton-like waves in (3D) experiments of weakly magnetized plasmas, in contrast to our findings of suppression of solitons in 2D \citep{Borghesi2002,Sircombe2005,Sarri2010,Henri2011}.
There have also been in-situ space measurements of solitary structures in the magnetospheres of solar planets exhibiting shapes similar to Langmuir solitons \citep{Pickett2015,Pisa2021}.
However, after a detailed consideration of these findings, we cannot decisively say that these experimentally and observationally detected waves are soliton-like cavitons or rather a special kind of ``linear'' Langmuir waves with similar properties.
We have shown in Figs.~\ref{fig2} and \ref{fig3}, that the solitons and Langmuir/L-mode waves can be identified by their specific features in the $\omega - k_\parallel$ dispersion diagram. 
These were, however, not provided for the mentioned laboratory and planetary observations.

\subsection{Implications on the pulsar radio emission}
Although the solitons are known to collapse and disappear, they are nevertheless often considered as sources of pulsar radiation because the plasma dispersion properties of extremely magnetized neutron star magnetospheres are 
conjectured to be well described by the 1D approach.

As the solitons do not appear in 2D because of the existence of additional degrees of freedom for the particles, their formation therefore also is not expected in 3D. 
When superluminal solitons cannot form for any reasonable set of plasma parameters, then solitons cannot be considered as an essential element of coherent curvature or any other emission mechanism in the inner neutron star magnetospheres.
Nevertheless, our findings cannot absolutely exclude the possibility of soliton formation via other plasma kinetic instabilities, and as we have discussed before, the 3D consideration may, however, further aggravate the described formation difficulties.

The absence of solitons does not constrain other possible radiation mechanisms of the streaming instability.
The streaming instability still forms strong local plasma charge variations that we found to be associated with the electrostatic waves.
Because the plasma charge density follows the positions of the electric field profile in Fig.~\ref{fig5}(d)--(f), other strong plasma waves might also lead to coherent curvature radiation.

Only point-like plasma charge concentrations have been considered in the past for predictions of the radio emission properties of solitons.
Realistic electric charge structures are, however, far from the point-source approximation usually considered in analytical calculations \citep{Melikidze2000}. 
The shapes and sizes of the charge structures are important as they significantly influence the observable radio emission patterns \citet{Benacek2023}.
The structures that formed along the perpendicular direction to the magnetic field, as shown in Fig.~\ref{fig5}, can in principle be expected to form strong field aligned curvature radiation patterns as individual radiating elements of the charge distribution may add up positively in the field direction. 
That alignment will then also enhance the observable radiative flux density. 
A more detailed knowledge of the spatial structure of solitons would allow a more self-consistent description of their distinctive radio emission properties, especially the directivity, frequency spectrum, and coherency of the produced waves.

 The relativistic plasma emission, linear acceleration emission, or the free electron maser were also suggested as alternatives to coherent curvature emission in pulsars \citep{Melrose2020a}.
The linear acceleration emission is however not effective enough because its emission power was found to be too low {($\approx$ $10^{16}$\,erg$\cdot$s$^{-1}$) to explain the total pulsar radio power ($10^{25}$--$10^{29}$\,erg$\cdot$s$^{-1}$) \citep{Benacek2023}.}
The specific radio emission properties of the other mechanisms are not well known because they are strongly nonlinear wave--particle processes.
Strongly overlapped bunches have, for example, a strong $E_\mathrm{y}$ component with a group velocity close to zero, oscillating at frequencies close to zero.
If relativistic plasma particles travel through this wave along a magnetic field, the particles will oscillate synchronously and produce radiation by the free electron maser/laser effect \citep{Lyutikov2020,Lyutikov2021a}. The amplification of free electron maser waves is, nonetheless, constrained by the curvature of the trajectory and the dissipation timescale, and it is still far from certain that it may be responsible for pulsar radio emission.

\subsection{Extrapolation of results to 3D and weaker magnetic fields}

It is an open question how the instability may evolve in 3D space and in a weaker magnetic field which may only be solved by fully fledged 3D simulations.
In 2D, the wave energy decrease can be understood as energy flowing from the superluminal wave ``volume'' (a region where most of the soliton electrostatic energy is located) to a surrounding plasma through a surface of the volume.
However, in 3D, the ratio between the surface and volume is even larger than in 2D, and the decrease rate can be expected to be higher.

Our simulations are consistent with magnetic fields $\gtrsim$$10^8$\,G, which is valid up to $\sim$$22\,R_\mathrm{ns}$ for neutron stars with a field intensity of $10^{12}$\,G.
Nevertheless, for millisecond pulsars, the condition may only be fulfilled at heights smaller than one star radius above the stellar surface.
 The plasma particles in weaker magnetic fields have, however, higher pitch angles w.r.t. the magnetic field, which acts against the soliton formation.

\citet{Bret2010} found that the streaming instability evolves in a significantly different manner in 2D compared to 1D for a weakly-relativistic, cold, electron--ion plasma with weak magnetic fields.
They found that oblique modes appear to grow and filamentation instabilities also grow and produce micro-currents in the perpendicular direction, and particles can be accelerated by the Weibel instability \citep{Nishikawa2003,Bret2006,Bret2008,Bret2010b}.
In addition, they estimated a typical perpendicular size of the modulated plasma in 3D simulations, finding a value $\sim$$10 d_{e}$.
In our 2D simulations, the typical perpendicular size of the L-mode waves is on the order of $\sim$$ 100 d_{e}$ for high Lorentz factors.


\section{Conclusions}
\label{sec:conclusion}
Streaming instabilities in the magnetospheres of neutron stars were theoretically proposed and later seen to form nonlinear caviton/soliton-like waves in various numerical 1D simulations. 
The instability may be formed because of plasma bunches overlapping in phase space, a drift between electrons and positrons, or by bunches propagating through a background plasma and the 1D approximation has, therefore, been assumed to be valid and useful as the approach could describe an instability evolution in a strongly magnetized plasma.
As solitons may host a large electric charge, they can be exploited to explain the intense pulsar radiation by, for example, the coherent curvature mechanism.

In this paper, we compared the 1D and 2D PIC simulations of streaming instabilities, using simulations for three typical parameter scenarios of the streaming instability based on analytical and numerical investigations that were known to form solitons in 1D.
We did not find any indication that solitons could be formed in 2D for those or any different parameter set under consideration. 
It appears that finding no solitons to form in 2D is because of plasma effects perpendicular to the magnetic field suppress the soliton formation.

The absence of the solitons generated by the analyzed relativistic streaming instability in a more realistic dimensionality (2D vs 1D) in pulsar star magnetospheres implies that the coherent radiation of pulsars, magnetars, and fast radio bursts is not generated by this kind of solitons that are known to exist only as a consequence of the necessary simplifications in 1-D simulations. 
Unless, a proven alternative mechanism for their creation can be established.

Whether other plasma modes can radiate by the coherent curvature emission mechanism or the streaming instability may radiate by other coherent radiation mechanisms, however, remains as an open question.
Detailed coherent radiation properties of the relativistic streaming instability in 2D will be studied in a follow-up paper.

\begin{acknowledgements}
We acknowledge George I. Melikidze for discussions of analytical solutions of nonlinear plasma wave equations.
We also acknowledge the support by the German Science Foundation (DFG) projects MU~4255, BU 777-16-1, BU~777-17-1, and BE~7886/2-1. 
We acknowledge the developers of the ACRONYM code (Verein zur F\"orderung kinetischer Plasmasimulationen e.V.).
The authors gratefully acknowledge the Gauss Centre for Supercomputing e.V. (\url{www.gauss-centre.eu}) for partially funding this project by providing computing time on the GCS Supercomputer SuperMUC-NG at Leibniz Supercomputing Centre (www.lrz.de), projects pn73ne.
\end{acknowledgements}

%
%

\begin{appendix}

\end{appendix}

\bibliographystyle{aa}
\bibliography{references}

\end{document}